# The application of the NeXus data format to ISIS muon data

S.P. Cottrell, D. Flannery, G. Porter, A.D. Hillier and P.J.C. King

*ISIS Facility, Rutherford Appleton Laboratory, Chilton, Didcot,*
*Oxon, OX11 0QX, United Kingdom*

**Abstract**

Although originally designed by and for the use of the Neutron and X-ray communities, the flexibility of the NeXus data format makes it equally suitable for storing data gathered from μSR experiments. Furthermore, its use should open up the possibility of sharing software beyond the immediate muon community; giving access to the many tools that are already in existence for manipulating NeXus and HDF based files. This paper explores the development of the NeXus format for storing ISIS muon data and the associated experimental conditions. The design of an Instrument Definition for the ISIS muon instruments is described and the implementation of an application to translate the present ISIS raw data format to the NeXus format presented. The development of a common muon data format is a topic of active discussion within the muon community; with this in mind, the suitability of NeXus and, in particular, the ISIS Instrument Definition for general application is considered. A number of applications that have been developed to exploit the flexibility of the ISIS muon NeXus file format are described and a scheme for enabling web based browsing and analysis of experiment data is introduced.

# 1. Introduction

Muons play an increasingly important role in the study of the structure and dynamics of materials, with a diverse range of applications in Physics and Chemistry, often with muon spectroscopy complementing other major probes such as neutrons, X-rays or magnetic resonance. The μSR technique [1] involves the implantation of spin polarized muons into a material. These muons are unstable and decay with a lifetime of approximately 2.2 μs, emitting a positron preferentially in the direction of the muon spin polarization. The time evolution of the muon polarization can be monitored by detecting these decay positrons, thereby obtaining information on the muons' local environment and behaviour. Positrons are typically detected using an array of plastic scintillators placed around the sample position, their precise configuration depending on the type of μSR experiment being carried out. The counts measured in each detector are histogrammed as a function of time using a time to digital converter (TDC) and stored in a raw data file; these histograms are later combined in software to form the signal used during data analysis.

Collection and storage of ISIS muon data is currently handled by computers running the VMS operating system, with raw data being held as records in a binary file. Here, an independent knowledge of their position is essential for an understanding of the file content. At present, each of the four major muon sources operating worldwide write data in a similar manner, but using a unique record format, making it difficult for scientists to combine data collected at the different facilities; it is now acknowledged [2, 3] that a common data format would be of great benefit to the μSR community. In addition to simplifying the exchange of data between facilities, a common format would also facilitate the sharing of analysis and visualization software and help cooperation in future program development. This paper explores the use of NeXus [4, 5] as a data format for storing ISIS muon data files and considers its more general suitability as a common μSR file format for data storage. An additional motivation for this work is that NeXus is the intended format for future storage of ISIS neutron data, and there are benefits within the facility in having a muon format that is familiar to neutron users.

The design of an Instrument Definition for writing NeXus files for the ISIS muon instruments is described and the implementation of a program (CONVERT_NEXUS) to convert the present ISIS binary data format to the NeXus format discussed. The extension of the ISIS Instrument Definition for use at other μSR facilities is also considered. A number of applications that have been developed to exploit the flexibility of the ISIS muon NeXus file format are described and a scheme for enabling web based browsing and analysis of experiment data is introduced.

**2. The ISIS muon Instrument Definition**

The ISIS muon Instrument Definition provides a complete representation of the structure and content of the ISIS muon NeXus files. In its design, advantage was taken of the flexibility of the NeXus format to store a complete description of the experiment conditions along with the data, additional details that have been incorporated with only a small (10–15%) increase in the overall file size when compared to the present ISIS binary data format. NeXus files therefore contain sufficient information to understand the experimental setup and subsequently carry out a full analysis. A complete description of the ISIS muon Instrument Definition is provided in Ref. 6, however key features are as follows:

- Data histograms are stored individually, enabling the detectors to be regrouped as required during analysis. The detector grouping most appropriate for the spectrometer and experiment is stored in the NeXus file. All the parameters required to interpret the histograms (such as histogram resolution, time zero and the first and last good histogram data bins) are stored in the file.
- The NeXus file format makes provision for including arrays of logged information within the file, together with the time at which each value was measured. This facility has been used to include a record of the sample environment temperature, measured at fixed time intervals throughout the experiment, and also the total number of detected positrons measured over fixed time intervals (usually 100 seconds at ISIS).
- Details of the sample environment equipment and a vector describing the orientation of the instrument magnetic field are included. Sufficient information (detector positions and solid angles) is contained in the file to enable a rudimentary model of the instrument detector arrays to be constructed.
- Dead-time information corresponding to each detector on the instrument is stored in the file. This facility ensures that values appropriate to the period when the data was collected are always available, enabling an accurate correction to be applied during analysis.

Many experiments carried out at ISIS make use of additional equipment to study, for example, the effects of electric fields, radio-frequency excitation or illumination on the sample. Storing additional parameters to describe these experiments is clearly important; however it is difficult for a file format to anticipate all of the many different physical quantities that might need to be saved. In the past, the free format comment field has been used to store additional information; this approach, however, is far from ideal as space for additional information is limited and it is difficult for data analysis programs to automatically extract and use the various parameters. The extensible nature of the NeXus data format makes storage of additional information simple, as parameters can easily be inserted into the tree structure as required without affecting the ability of existing analysis software to read the data files.

**3. Towards a common NeXus Instrument Definition for μSR**

A preliminary version of the ISIS muon Instrument Definition was presented at the NeXus workshop held at PSI, Switzerland, in March 2001. The opportunity was taken to discuss the content of NeXus Instrument Definitions appropriate for muon instruments with the other members of the μSR community present at the meeting. It was concluded that it would be difficult to devise a single Instrument Definition that would be appropriate to all μSR spectrometers, but instead advantage should be taken of the extensible nature of the data format to define a subset of essential element names, units, coordinate systems and data types that would be present in all muon NeXus files. It was recognised, however, that laboratories would then extend on these core elements, although it was envisaged that certain elements that are unnecessary for every type of μSR experiment but likely to be common in a particular field of μSR could be standardised.

Contained within the NeXus file is a standard element name that defines the type of μSR experiment carried out (for the present ISIS instruments this is always set to 'muonTD', denoting Time Differential experiments). Therefore, given a limited degree of standardisation between facilities, NeXus-aware analysis software might automatically discover the type of experiment carried out, extract basic information relevant to this experiment and then carry through a preliminary analysis of the data, irrespective of the institute where the data was written. By then reading the institute name (also stored in the NeXus file) and correlating this with the appropriate NeXus Instrument Definition, the analysis software could then read the additional information defined by the facility.

**4. Using NeXus files at ISIS**

The adoption of any new data file format at a user facility is a major undertaking. Therefore, as a first step in the process of evaluating NeXus, it was decided to create an application (CONVERT_NEXUS) to run on the VMS based ISIS data acquisition computers, to translate existing ISIS muon binary data files into NeXus files that conform to the ISIS muon Instrument Definition. In this way the user retains the choice of using either the original file format or moving to NeXus.

In use, CONVERT_NEXUS brings numerous advantages beyond simple file conversion. Of particular value is its ability to combine into the NeXus file elements which have previously resided in numerous separate run log files. A software engineering context diagram (Fig. 1) shows the interactions between the terminators and the CONVERT_NEXUS application; key features are as follows:

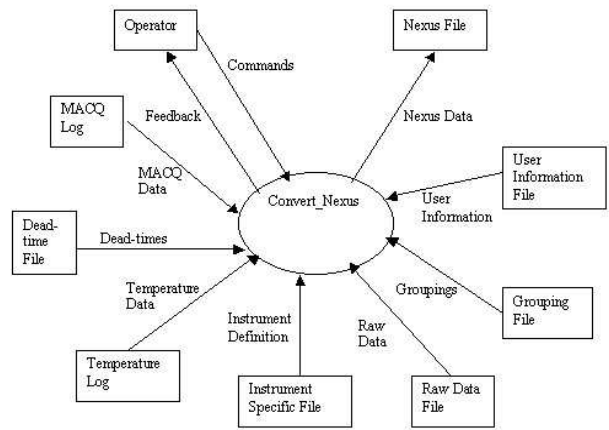

Figure 1: A Software Engineering Context diagram denoting the interactions between the terminators and the CONVERT_NEXUS application.

- Temperature and data acquisition log files are included in the NeXus file. Each logged value is time stamped and temperature logs that were previously split by the data acquisition system over several different files are integrated.
- Information about the instrument collecting the data is incorporated by reading an Instrument Specific File; this currently contains the detector positions and solid angles. If available, a file containing the detector dead-times for the instrument is included, as is an appropriate detector grouping file.

- An exact value for time zero (the time from the start of data acquisition to the centre of the muon pulse) is read and included in the file.
- A mechanism is available for users to include additional information (read from a User Information File) to describe unique experimental features.
- The conversion program has been designed to operate either interactively or silently with only fatal error messages being displayed. This latter mode has the advantage that CONVERT_NEXUS can be run from within a VMS script to automate the file conversion process.

Complementary to the CONVERT_NEXUS application, NeXus reader subroutines have been programmed in Fortran 77 and C to enable information stored according to the ISIS NeXus muon Instrument Definition to be read by user applications. These routines have been compiled and tested under the Microsoft Windows, Linux and VMS operating systems, and are freely available [6].

**5. Muon NeXus Applications**

Already a number of applications such as UDA (a µSR data analysis program commonly used at ISIS) have been adapted to directly read NeXus files, and a utility program written to plot temperature and event log data as a function of time, side-by-side for comparison (example output is shown in Fig. 2).

A major development that is currently underway centers on the application of the OpenGenie data visualization package [7] to µSR data. A particular motivation for this work is that OpenGenie will form the basis of the new Instrument Control Program (ICP) that will come into use over the next few years as ISIS moves to a PC based data acquisition system. OpenGenie, with its built in HDF libraries, can easily read and write muon data directly in the NeXus file format, thus allowing 'on the fly' data analysis and bringing the possibility of collecting data to a specific statistical accuracy. With a built in programming language and the ability to link to external subroutines, the package is extremely flexible and can be readily extended to offer the unique facilities required by muon spectroscopy. Currently, a suite of subroutines has been developed for instrument debugging, in which individual detectors can viewed and analysed, run details examined and log parameters plotted. Work is continuing to code the basic data reduction and analysis routines required for µSR analysis; these make use of the internal least squares fitting routines already available within OpenGenie.

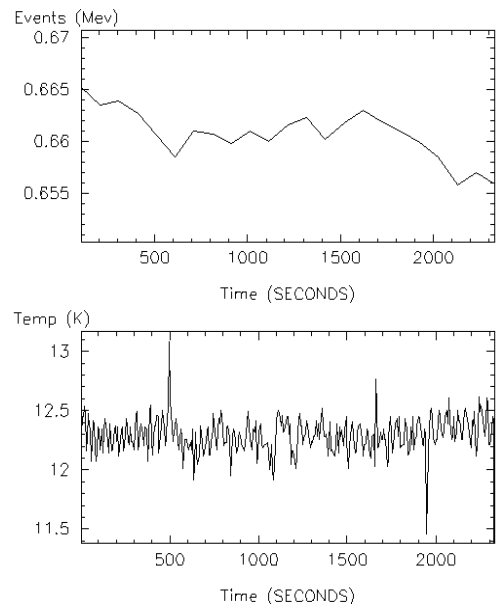

Figure 2: Temperature and event log data

**6. Web based browsing and analysis of NeXus data files**

A project is now in progress to provide browsing, plotting and, potentially, simple analysis of muon NeXus data files across the web. The original aim of this work was to enable scientists to view a snapshot of their data before transferring what are often large datasets between institutions. With ever increasing processing power available in the client computer, however, the broader goal of providing full on-line data reduction, analysis and visualisation now appears to be a possibility. It is clear that there is considerable interest in having this type of access to datasets collected at central facilities.

Within the µSR community, Brewer *et al* [2] have developed a Java based scheme for browsing and analysis of data files collected at the Canadian muon source TRIUMF (although in this case data processing was carried out by the server). Also, a scheme for network browsing of NeXus data files has been investigated by Könnecke [8].

Written in Java, our project makes use of the Servlet and Remote Method Invocation technologies for seamless remote communication between client and server, as depicted in Fig. 3. The benefit of employing Servlets is that security, firewall and proxy server issues are eliminated, thereby making the system fully Internet deployable. At present, the principle role of the server is to translate the NeXus file into a form suitable for both Internet transfer and subsequent manipulation by Java. The Extensible Markup Language (XML) was chosen as a suitable representation, since this appears to be emerging as a standard tool for writing NeXus data files in ASCII format with the file structure preserved, and additionally offers the possibility of validating the data files according to the appropriate Instrument Definition File [5]. When a request for a NeXus file is made, the server triggers a process to recursively convert the entire NeXus file to XML 'on the fly'. During this conversion, the XML is validated using a DTD (generated from a meta-DTD produced using the recently implemented NXtoDTD utility [5]) to ensure the files are standard-conforming; in practice the entire dataset is returned to the client application with non-standard sections highlighted. Once in the appropriate format, the data is available for processing and manipulation by the client.

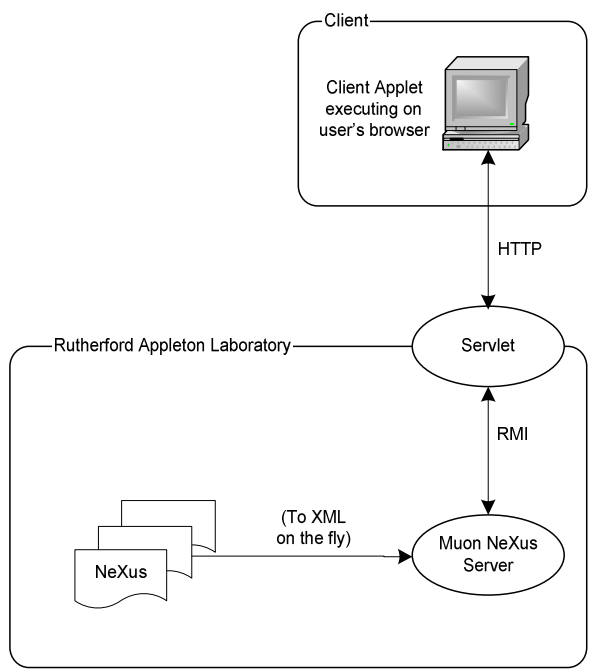

Figure 3: High level design of client-server model for web based browsing of muon data files.

The client consists of a lightweight Java Applet embedded in a web page. Applets are particularly useful for this application as they support complex and powerful user interfaces, enabling structured access to the data for histogram plotting and visualization. Additionally, Applets are immediately available to anyone with an Internet connection and Java enabled web browser (the code is platform independent), and their use removes the need for client software installation: if a change in the software occurs, the new version of the software is immediately and automatically available to all users.

In the present implementation the complete dataset is transferred to the client, and therefore the data transfer rate on the network plays a critical role in the application performance. However, for its intended use as a method of access for NeXus muon data files containing relatively small datasets, this does not represent a significant limitation.

**7. Conclusion**

It is already clear from our current work that there are considerable benefits for µSR users in adopting a structured and self describing format such as NeXus that is already used and supported by a wider community of scientists. A key advantage of the NeXus format is that information can be added to the data file without affecting the ability of existing subroutines to read (a subset of) the extended data files. This feature has enabled us to evolve the Instrument Definition into its present form and, more importantly, allows experiment specific information to be added to the data file to describe novel apparatus. Hopefully, this flexibility will encourage the other muon facilities to develop suitable Instrument Definitions and adopt NeXus in some way, perhaps initially as an intermediate data format with appropriate file converters. It is important, however, that dialog between the facilities occurs at an early stage in order that some standardisation of the Instrument Definitions can be agreed; without this, the full benefits offered by the NeXus data format may not be realised. Discussions with users of the NeXus format in the wider neutron and X-ray communities would also be valuable as it is important that muon Instrument Definition Files follow agreed standards as closely as possible.

There are many possibilities for the development of NeXus software, and work on applications will continue in anticipation that NeXus offers a viable approach for evolving a common muon file format that will enable µSR users to easily share data and analysis programs. Of current interest is our project to develop a framework that will allow NeXus files to be served over the web to a Java Applet installed in the clients web browser – a configuration that offers users a potentially truly cross-platform solution for data manipulation. Work is continuing with the coding of a sophisticated Java Applet to provide all the common subroutines necessary for reduction, analysis and visualisation of µSR data. In the course of this project we anticipate carrying out a detailed evaluation of the performance of the framework in terms of the data transfer rate and the computational performance of the client Applet. Depending on the outcome of this work, the possibility of deploying complex data analysis routines for execution on the server will be explored (where results would be returned to the client only for visualization), and an extension to the framework may be considered such that partial datasets can be transferred to reduce the load when large datasets are being handled.

**Acknowledgements**

We would like to thank Dr T.M. Riseman for useful discussions throughout this project.